# Estimating Unpropped Fracture Conductivity and Fracture Compliance from Diagnostic Fracture Injection Tests[1] [2]

HanYi Wang and Mukul M. Sharma, The University of Texas at Austin

## Abstract

A new method is proposed to estimate the compliance and conductivity of induced unpropped fractures as a function of the effective stress acting on the fracture from DFIT data. A hydraulic fracture's resistance to displacement and closure is described by its compliance (or stiffness). Fracture compliance is closely related to the elastic, failure and hydraulic properties of the rock. Quantifying fracture compliance and fracture conductivity under in-situ conditions is crucial in many earth science and engineering applications but very difficult to achieve. Even though laboratory experiments are often used to measure fracture compliance and conductivity, the measurement results are strongly influenced by how the fracture is created, the specific rock sample obtained and the degree to which it is preserved. As such the results may not be representative of field scale fractures

Over the past two decades, Diagnostic Fracture Injection Tests (DFIT) has evolved into a commonly used and reliable technique to obtain in-situ stresses, fluid leak-off parameters and formation permeability. The pressure decline response across the entire duration of a DFIT test reflects the process of fracture closure and reservoir flow capacity. As such it is possible to use this data to quantify changes in fracture conductivity as a function of stress. In this paper we present a single, coherent mathematical framework to accomplish this. We show how each factor impacts the pressure decline response and the effects of previous overlooked coupled mechanisms are examined and discussed. Synthetic and field case studies are presented to illustrate the method. Most importantly, a new specialized plot (normalized system stiffness plot) is proposed, which not only provides clear evidence of the existence of a residual fracture width as a fracture is closing during a DFIT, but also allows us to estimate fracture compliance (or stiffness) evolution and fracture un-propped conductivity using only DFIT pressure and time data based on a time-convolution solution. It is recommended that the normalized system stiffness plot be used as a standard practice to complement the G-function or square root of time plot because it provides very valuable information on the properties of fracture surface roughness at a field-scale, information that cannot be obtained by any other means.

## Introduction

Fractures are ubiquitous in the subsurface. Small-scale fractures like cracks, fissures and large-scale fractures like joints and faults are key structures that determine the mechanical resistance and the fluid transport properties of rocks. The flow and mechanical properties of these fractures are controlled by in-situ stress and its compliance or stiffness (fracture stiffness is the reciprocal of fracture compliance and these terms will be used interchangeably throughout this article), which are primarily controlled by the rock mineralogy and the fracture surface roughness (Hopkins et al., 1987; Pyrak-Nolte and Nolte 2016). In unconventional reservoirs, the permeability of the stimulated reservoir volume (SRV) created around hydraulic fractures is dominated by the properties of the induced unpropped (IU) fractures (Sharma and Manchanda 2015; Wang 2017). Characterizing fracture compliance is, therefore, crucial in addressing not only the mechanical, hydraulic, and transport properties of a fracture in the subsurface but also well productivity and ultimate recovery of wells in low permeability rocks. Other applications where fracture compliance plays a central role include, fault zone studies (Scholz 2002), underground $CO_2$ sequestration (Iding and Ringrose 2009), nuclear waste repositories (Witherspoon 2004) and geothermal energy exploitation (Evans et al. 1992).

The average fracture width $\overline{w}_f$ and fracturing net pressure $P_{net}$ is related by the fracture stiffness $S_f$ for an open fracture:

$$P_{net} = S_f \overline{w}_f \qquad (1)$$

Physically, fracture stiffness defines how compressible the fracture is. Apply the theory of linear elasticity, a closed-form solution to the displacement and stress distribution in the interior of an elastic solid by the opening of an internal crack under the action of pressure applied can be obtained. **Table 1** shows the fracture stiffness for different fracture geometries.

| Fracture Geometry | PKN | KGD | Radial |
|---|---|---|---|
| $S_f$ | $\dfrac{2E'}{\pi h_f}$ | $\dfrac{E'}{\pi x_f}$ | $\dfrac{3\pi E'}{16 R_f}$ |

**Table 1-Fracture stiffness expressions for different fracture geometry models**

where $E'$ is the plane strain Young's modulus and can be calculated using Young's Modulus, $E$, and Poisson's Ratio, $v$:

$$E' = \frac{E}{1-v^2} \qquad (2)$$

The fracture stiffness calculated from the Table 1 assumes that the fracture closes completely and instantaneously when the fluid pressure in the fracture reaches the closure pressure (so fracture stiffness is only related to fracture geometry and rock properties). In recent years, advanced numerical models (Bryant et al. 2015; Sesetty and Ghassemi 2015; Wang 2015; Wang 2016; Wang et al. 2016) have been proposed to model hydraulic fracturing process assuming smooth crack surfaces. However, during the fracture closure process, the fracture will gradually close and the stress acting normal to the fracture plane results in fracture apertures approaching the scale of the surface roughness, and fracture surfaces can no longer be treated as perfectly smooth. Thus fracture stiffness is no longer a constant value once the fracture starts to close on asperities and rough walls.

Fracture faces are never smooth and instead have rough-walled structures. van Dam et al. (2000) presented scaled laboratory experiments on hydraulic fracture closure behavior. They observed up to a 15% residual aperture (compared to the maximum aperture during fracture propagation) long after shut-in. Fredd et al. (2000) demonstrated that fracture surface asperities can provide a residual fracture width and sufficient conductivity in the absence of proppants. Using sandstone cores from the East Texas Cotton Valley formation, sheared fracture surface asperities that had an average height of about 2.286 mm were observed. Warpinski et al. (1993) reported hydraulic fracture surface asperities of about 1.016 mm and 4.064 mm for nearly homogeneous sandstones and sands with coal and clay-rich bedding planes, respectively. Sakaguchi et al. (2006) created a tensile fracture on large rock blocks and measured the asperity height and distribution. Their work shows that the fracture surfaces can be assumed to be a fractal object while most of the asperities fall within a size range of 1-2 mm. Wells and Davatzes (2015) conducted topographic measurements on dilated fractures from core samples and found that the asperity heights range from hundreds to thousands of micrometers. Bhide et al. (2014) created X-ray microtomographic images from shear-induced fractures and the roughness values obtained varied from 1.8 to 1.95 mm along the length of rock samples. Zou et al. (2015) conducted experiments on 20 fractured shale samples and found the average asperity height to be 1.88 mm. Field measurements (Warpinski et al. 2002) using a down-hole tiltmeter array indicated that the fracture closure process is a smooth, continuous one which often leaves 20%-30% residual fracture width, regardless of whether the injection fluid is water, linear-gel or cross-linked-gel.

The degree of contact between fracture faces controls both fracture mechanical properties and its conductivity, and our ability to quantify fracture conductivity under different in-situ stress conditions is crucial in optimizing hydraulic fracture design (e.g., maximize total fracture surface area vs. maximize propped fracture length) and field development. In this study, we propose a new approach to estimate fracture compliance and unpropped fracture conductivity from diagnostic fracture injection test (DFIT) data. The structure of this article is as follows. First, we show why fracture compliance can change during fracture closure and how it is related to fracture surface roughness and fracture residual conductivity. Next, we presented a new semi-analytical DFIT model that models both before-closure and after-closure data and integrates them seamlessly by accounting for both variable fracture compliance and fracture pressure dependent leak-off. Then, synthetic and field cases are analyzed to validate our approach and the virtue of a new specialized plot. Finally, conclusions and discussions are presented. The mathematical derivations of our semi-analytical DFIT model and time-convolution solution are summarized in the Appendix.

## Variable Fracture Compliance

There are basically two main causes that lead to the continuously changing fracture compliance during fracture closure. The first is stress contrast in the different layers penetrated by the fracture. In this case, fracture will close first in the zones where the minimum in-situ stress is highest. This alters the overall fracture stiffness during the closure process. The second cause of variable fracture compliance is the presence of fracture surface asperities and roughness. As the fracture closes on asperities progressively from its edges to the center, the overall fracture stiffness is determined by both the closed portion and open portion of the fracture.

The microscopic measurement and modeling of surface roughness and mechanical properties of asperities can often be up-scaled to macroscopic contact laws that relate fracture width to the associated contact stress. Willis-Richards et al. (1996) proposed a contact law to relate fracture width and the net closure stress for fractured rocks, based on the work of Barton et al. (1985):

$$\sigma_c = \frac{\sigma_{ref}}{9}\left(\frac{w_0}{w_f} - 1\right) for \ w_f \leq w_0 \qquad (3)$$

where $w_f$ is the fracture aperture and, $w_0$ is the contact width, which represents the fracture aperture when the contact normal stress is equal to zero, $\sigma_c$ is the contact normal stress on the fracture, and $\sigma_{ref}$ is a contact reference stress, which denotes the effective normal stress at which the aperture is reduced by 90%. The contact width $w_0$ is determined by the tallest asperities, and the strength, spatial and height distribution of asperities are reflected in the contact reference stress $\sigma_{ref}$ (e.g., if the tallest asperities on two fracture samples are the same, then they should have the same $w_0$, but the one with a higher median asperity height, Young's modulus or yield stress will have higher value of $\sigma_{ref}$, provided other properties are the same).

Both $w_0$ and $\sigma_{ref}$ can be obtained from laboratory experiments with fractured rock samples, with controlled measurements of $\sigma_c$ and $w_f$. With known fracture geometry, rock properties and surface roughness (represented by contact parameters $w_0$ and $\sigma_{ref}$), the question now is how to estimate fracture stiffness (or compliance) as a function of pressure, since it continuously changes during fracture closure. McClure et al. (2016) presented a numerical model that incorporates this contact law (i.e., Eq.(3)) into hydraulic fracturing simulation. Wang and Sharma (2017a) proposed an integral transform method and general algorithms to model the dynamic behavior of hydraulic fracture closure on rough fracture surfaces and asperities, using linear elastic solutions that are coupled with a contact law for three different fracture models (PKN, KGD and radial fracture geometry). Given the fracture geometry, rock properties, contact parameters and minimum principal stress, their approach can predict the evolution of fracture aperture profile, total fracture volume, non-uniform contact stress distribution and fracture stiffness as fracturing pressure declines. Their study reveals that fracture does not close like parallel plates, but rather it closes on asperities progressively from edges to the center. Wang et al. (2017) presented an improved model for fracture closure based on superposition principles. Their model can simulate large scale fracture closure behavior with layer stress contrast in an efficient manner. **Fig.1** shows the simulated fracture volume and stiffness evolution for a PKN fracture geometry, assuming a Young's modulus of 20 GPa, Poisson's ratio of 0.25, $w_0$ of 2 mm, $\sigma_{ref}$ of 5 MPa with 10 m fracture height, 35 MPa minimum in-situ stress, together with a tiltmeter measurement (measure the deformation during fracture closure and is proportional to average fracture width and residual fracture volume, its slope is proportional to fracture compliance) from GRI/DOE M-site project. As can be seen, if fracture surface is completely smooth, the fracture stiffness remains constant until fluid pressure drops to minimum in-situ stress. This leads to a sudden increase in fracture stiffness when all fracture surfaces come into contact simultaneously. However, because of surface asperities and rough fracture walls, the fracture stiffness only remains constant at high fracture pressures. The stiffness starts to increase long before the pressure in the fracture declines to the minimum in-situ stress. An abrupt change in fracture stiffness can lead to abnormal spikes of pressure derivatives on diagnostic plots (Wang and Sharma 2017b), which is never observed in the field, so the fracture stiffness must change gradually during closure. van den Hoek (2005) also noted that a fracture has to shrink either in the height direction or length direction with variable fracture compliance to avoid unphysical spikes of pressure derivatives for waterflood-induced fracture closure. Both non-local fracture modeling results and tiltmeter measurements confirmed that fracture closure is a smooth and gradual process because of rough fracture walls and asperities, so it is difficult to accurately pick closure stress from tiltmeter measurements or even flow-back data (Wang and Sharma 2017a).

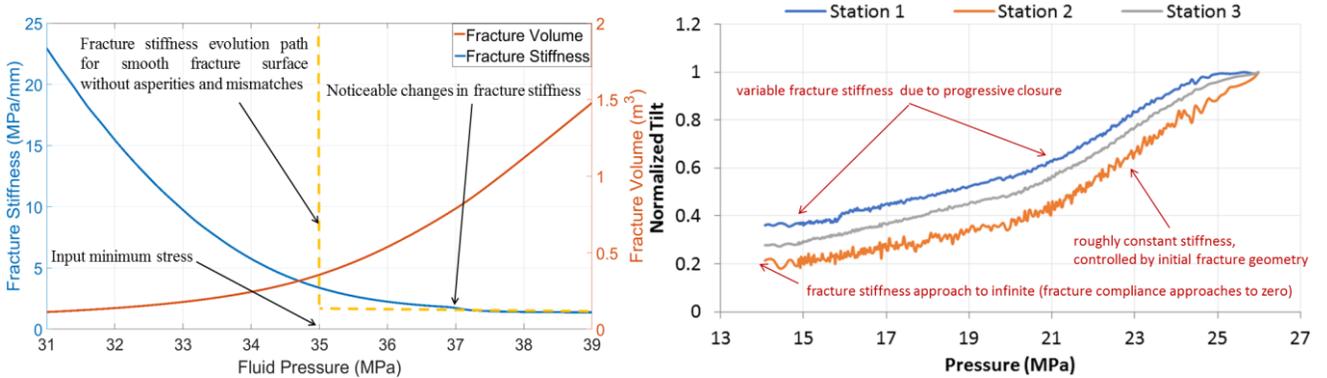

**Fig.1 Illustration of fracture stiffness evolution from modeling (left) and tiltmeter measurements (right) as pressure declines (modified from Wang and Sharma 2017a).**

## Relate Surface Roughness and Conductivity

Knowing the properties of surface roughness (represented by up-scaled contact width $w_0$ and contact reference stress $\sigma_{ref}$), and applying the non-local fracture closure model (Wang and Sharma 2017a; Wang et al 2017), the fracture width profile at any fluid pressure can be determined, regardless of whether the fracture is open, partially closed or completely closed (all asperities have come into contact with stress-dependent residual fracture width). Since the matrix permeability is low and there exists a linear relationship between the flow rate and differential pressure along the fracture during production, the fracture permeability $k_f$ can be calculated based on the cubic law (Watanabe et al., 2008):

$$k_f = \frac{w_f{}^2}{12} \tag{4}$$

The fracture conductivity $C_f$ is defined as the product of fracture permeability and fracture width:

$$C_f = k_f w_f \tag{5}$$

For a fracture with arbitrary fracture width distribution, as shown in **Fig.2**, the fracture cross section area that is perpendicular to fracture flow can be discretized into a number of sections. In the $i^{th}$ section, the cross-section area is $A_i$, the average fracture width is $\overline{w_{f(i)}}$ and the corresponding conductivity is $C_{f(i)}$. To get representative fracture conductivity over the entire fracture cross-section area, an arithmetic average needs to be applied:

$$C_f = \frac{\sum_{i=1}^{i=n} C_{f(i)} A_i}{\sum_{i=1}^{i=n} A_i} \tag{6}$$

Combining Eq.(4) and Eq.(5) into Eq.(6), and assuming the fracture is uniformly discretized in the y-direction, then the calculation of fracture conductivity can be simplified as:

$$C_f = \frac{1}{12} \frac{\sum_{i=1}^{i=n} \overline{w_{f(i)}}^4}{\sum_{i=1}^{i=n} \overline{w_{f(i)}}} \tag{7}$$

Because fracture geometry, rock mechanical properties and contact parameters uniquely determine the fracture width profile at a given fluid pressure, the fracture conductivity evolution during fracture closure can be estimated once the dynamics of fracture width profile is known. However, the question remains: how can we obtain representative values of contact width $w_0$ and contact reference stress $\sigma_{ref}$ for a field scale fracture, rather than from small scale laboratory experiments. Is it possible to obtain contact parameters from DFIT data? We will examine the impact of surface roughness on pressure response during fracture closure and the possibility of estimating contact parameters by DFIT data in later sections.

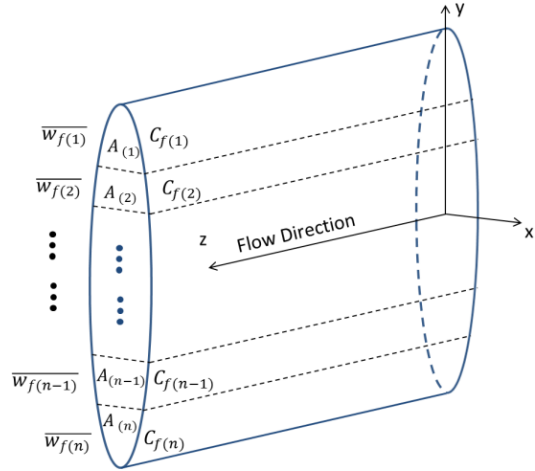

**Fig.2 Illustration of discretizing cross-section area of fracture flow.**

## Diagnostic Fracture Injection Test

Diagnostic Fracture Injection Tests, which have also been referred to as Injection-Falloff Tests, Fracture Calibration Tests, Mini-Frac Tests in the literature, involve pumping a fluid (typically water), at a constant rate for a short period of time, creating a relatively small hydraulic fracture before the well is shut in. The pressure transient data after shut-in is analyzed to obtain hydraulic fracturing parameters and reservoir properties. A typical pressure trend is qualitatively shown in **Fig.3**.

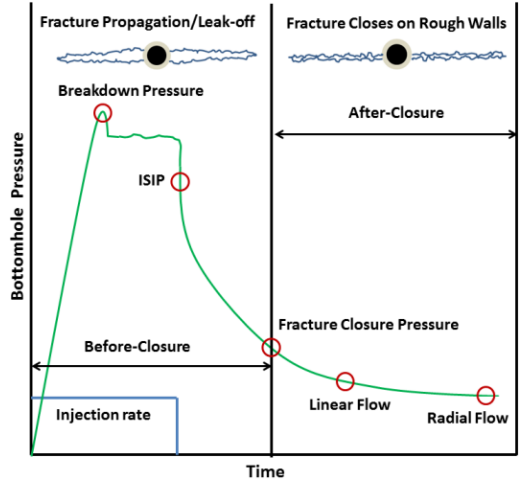

**Fig. 3 Diagram showing sequence of events observed in a DFIT**

The advent of fracturing pressure decline analysis was pioneered by the work of Nolte (1979 and 1986). With the assumptions of power-law fracture growth, negligible spurt loss, constant fracture surface area after shut-in and Carter's leak-off model, a remarkably simple and useful equation for the pressure decline can be obtained:

$$P_f(\Delta t_D) = \text{ISIP} - \frac{\pi r_p C_L S_f \sqrt{t_p}}{2} G(\Delta t_D) \qquad (8)$$

Here, ISIP is the instantaneous shut-in pressure at the end of pumping, $P_f$ is the fracture pressure at dimensionless time $\Delta t_D$. $t_p$ is the pump time. $r_p$ is the productive fracture ratio, which is the ratio of fracture surface area that is subject to leak-off to the total fracture surface area. $G$ is a dimensionless function. The productive fracture ratio $r_p$ is related to the heterogeneity in the permeability of rock layers the fracture has penetrated. In heterogeneous conventional reservoirs, a fracture can penetrate layers in which the permeability ranges from millidarcy (target layer) to nanodarcy (barrier layers), so $r_p$ can be less than 1 because the leak-off in barrier layers is negligible. In unconventional reservoirs $r_p$ should be close to 1, as evidenced by the high fluid efficiency and long closure time (the entire fracture surface is subject to leak-off with a small leak-off rate ). Eq.(8) forms the underlying basis of the G-function plot. Wang and Sharma (2017b) derived the solution to interpret the square root of time plot:

$$P_f(\Delta t) = \text{ISIP} - 4S_f(\text{ISIP} - P_0)\sqrt{\frac{k\phi c_t \Delta t}{\pi \mu_f}} \qquad (9)$$

where k is formation permeability, $\mu_f$ is fluid viscosity, $\phi$ is formation porosity, $c_t$ is total formation compressibility and $P_0$ is the initial reservoir pressure. Their study reveals that both G-function plot and square root of time plot are equivalent, and gives the same quantitative information with slightly difference in scales. Because both Eq.(8) and Eq.(9) have two distinct and important assumptions: (1) Carter's leak-off with constant fracture pressure boundary condition and, (2) fracture stiffness is assumed to be constant during fracture closure

To resolve these inconsistencies between the fundamental assumptions and reality, Wang and Sharma (2017b) presented a new DFIT model which accounted for fracture pressure dependent leak-off (FPDL) and variable fracture compliance. In this study, we use their work as a starting point and derive a new time-convolution solution for pressure transient behavior during a DFIT. A detailed derivation of the pressure decline solution is presented in the **Appendix**. The fracture pressure at the n[th] time interval, $P_{f,n}$, can be calculated explicitly using time-convolution:

$$P_{f,n} = ISIP - 4\sqrt{\frac{k\phi c_t}{\pi \mu_f}} \sum_{i=1}^{n-1} S_{f,i} \sum_{j=1}^{i} (P_{f,j} - P_{f,j-1})\left(\sqrt{\Delta t_i - \Delta t_{j-1}} - \sqrt{\Delta t_{i-1} - \Delta t_{j-1}}\right) \qquad n \geq 2 \qquad (10)$$

where $P_{f,0} = P_0$, $P_{f,1} = ISIP$, $\Delta t_1 = 0$. $S_{f,i}$ is the fracture stiffness that corresponds to fracture pressure $P_{f,i}$ at shut-in time $\Delta t_i$. To account for wellbore storage effects, fracture stiffness $S_f$ needs to be replaced by fracture-wellbore system stiffness $S_s$ as defined by Eq.(A10). It should be emphasized that Eq.(10) does not distinguish between before closure and after closure period, it is a global pressure transient model and the fracture closure process is implicitly reflected in $S_{f,i}$. The relationship between $S_f$ and $P_f$, such as the curves in Fig.1, can be obtained from non-local fracture closure modeling (Wang and Sharma 2017a; Wang et al. 2017), where the pressure-dependent fracture stiffness is calculated by inputting rock property, fracture geometry and surface roughness (represented by contact parameters $w_0$ and $\sigma_{ref}$ ).

By forward modeling, Eq.(10) gives us a clear indication of what factors control the pressure decline behavior, and this enables us to investigate the effects of fracture surface roughness on pressure response through a pressure-dependent fracture stiffness, and in turn, we can explore the possibility of extracting useful information about fracture compliance and conductivity based on a global analysis and history match of the DFIT data. By inverse modeling using just pressure and time data through Eq.(A12), we can estimate the relative changes of fracture or fracture-wellbore system stiffness, and infer pressure-dependent fracture conductivity qualitatively. In the following sections, sensitivity analysis will be presented with forward modeling approach, and both forward and inverse modeling will be discussed and compared for some field cases.

## Sensitivity Analysis of Factors Impact Pressure Decline Signature

In this section, Eq.(10) is used to investigate how different fracture geometry, contact parameters and reservoir properties impact the fracture compliance/stiffness evolution and pressure decline response during DFIT, so that we can differentiate the signatures on a pressure transient response that is mostly controlled by fracture surface roughness. Assuming a Base Case scenario where the input parameters are provided in **Table** 2.

| | |
|---|---|
| Fracture type | PKN |
| Fracture height | 10 m |
| Fracture length | 50 m |
| Contact width | 2 mm |
| Contact reference stress | 5 MPa |
| Pumping time | 5 min |
| ISIP | 40 MPa |
| Minimum in-situ stress | 35 MPa |
| Initial pore pressure | 20 MPa |
| Reservoir permeability | 0.0005 md |
| Young's modulus | 20 GPa |
| Total compressibility | 1.9e-3 MPa$^{-1}$ |
| Viscosity | 1 cP |
| Poisson's Ratio | 0.25 |
| Initial porosity | 0.03 |

**Table 2-Input parameters for Base Case scenario**

**Fig.4** shows the contact stress at different local fracture widths. As expected, when the fracture width is larger than the contact width $w_0$, the contact stress is zero. However, when the fracture width is smaller than the contact width, the contact stress and fracture width follows a hyperbolic relationship, as reflected by Eq.(3). **Fig.5** shows the corresponding fracture stiffness evolution based on the solutions of linear elasticity (Wang and Sharma 2017a; Wang et al. 2017), for different $w_0$ with the given fracture geometry and rock properties (provided in Table 2). The results indicate that as the fracture width decreases, the rough fracture walls will come into contact sooner if the contact width is larger, so the noticeable changes of fracture stiffness occur earlier when the contact width is larger. We can also observe that when the contact width is larger, the increase in fracture stiffness is more gradual and smooth. In the extreme case when $w_0$ is zero, as would be the case for perfectly smooth fracture surfaces, the fracture stiffness will increase abruptly to infinity as soon as the fluid pressure drops to the minimum in-situ stress (35 MPa in this case). This pressure-dependent fracture stiffness can be directly put into Eq.(10) to predict the fracture pressure decline for certain reservoir properties.

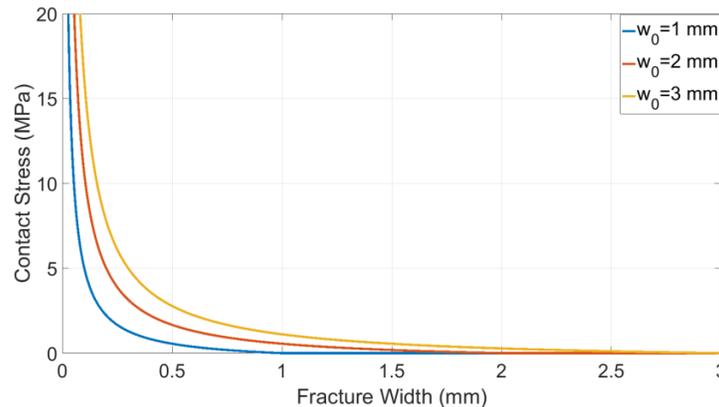

**Fig.4 The relationship between contact stress and fracture width for different $w_0$**

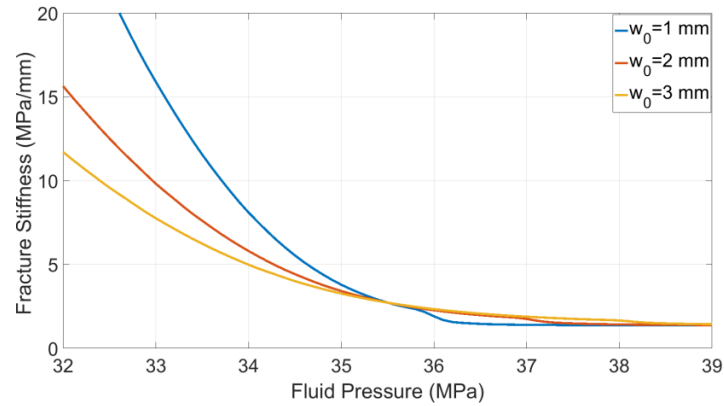

**Fig.5 Fracture stiffness evolution for different $w_0$**

**Fig.6** shows the fracturing pressure and its derivatives for different $w_0$ on G-function and square root of time plots. We can notice that the contact width impacts the pressure decline response significantly, because it alters the evolution of the fracture stiffness. Large contact width leads to a smooth pressure decline trend while small contact width leads to steep changes in the pressure decline rate and pressure derivatives. We can also infer that if the contact width is close to zero and all fracture walls come into contact simultaneously, then a sudden change in pressure decline rate and the pressure derivative spikes on both the G-function and square root of time plots. This is unrealistic and never observed in field cases. So the conventional assumption that a fracture closes on flat, smooth fracture surfaces where $w_0 = 0$ does not reflect reality. The most important observation is that at the beginning, pressure and its derivative are not impacted by $w_0$. This is because at higher pressure, the fracture walls are still wide open and the contact of asperities at the fracture edges has a negligible influence on the overall fracture stiffness, which is mainly controlled by the initial fracture geometry (as determined in Table 1). So the early straight line period can be used to constrain the fracture geometry on a G-function plot (use Eq.(8)) or square root of time plot (use Eq.(9)).

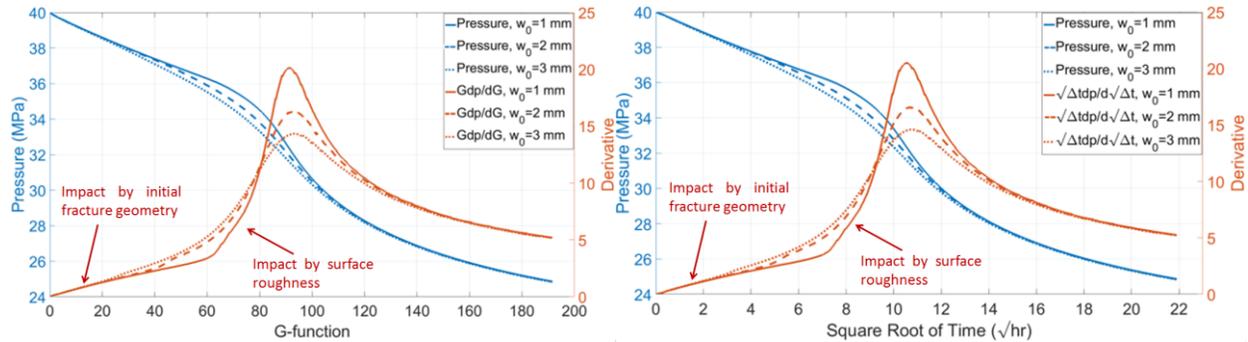

**Fig.6 Pressure decline response for different $w_0$**

**Fig.7** shows the fracture conductivity (1 um$^2$-cm $\approx$ 33 md-ft). as a function of increasing effective stress (far-field minimum in-situ stress minus fluid pressure inside the fracture) for different contact width $w_0$. As expected, the larger the $w_0$, the higher the fracture conductivity because the overall fracture residual aperture is larger. We can also see that when $w_0$ is smaller, the fracture conductivity is more sensitive to an increase in the effective stress.

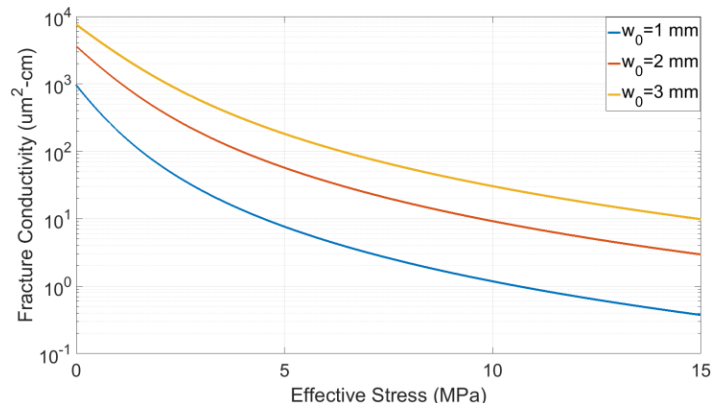

**Fig.7 Fracture conductivity evolution for different $w_0$**

Next, we examine how the contact reference stress affects the pressure decline response. **Fig.8** and **Fig.9** show the relationship between contact stress and fracture width for different contact reference stress and the corresponding fracture stiffness evolution at different fracturing pressure. For the same contact width, the higher the contact reference stress, the more rapid the increase of contact stress as the fracture width shrinks. Physically, the contact reference stress represents how hard and strong the fracture surface asperities are. The lower the contact reference stress, the more gradual the change in fracture stiffness as pressure declines. Even though the contact reference stress does not have much impact on the pressure at which the fracture stiffness starts to changes noticeably, it does impact the fracture stiffness evolution, as shown in Fig.9.

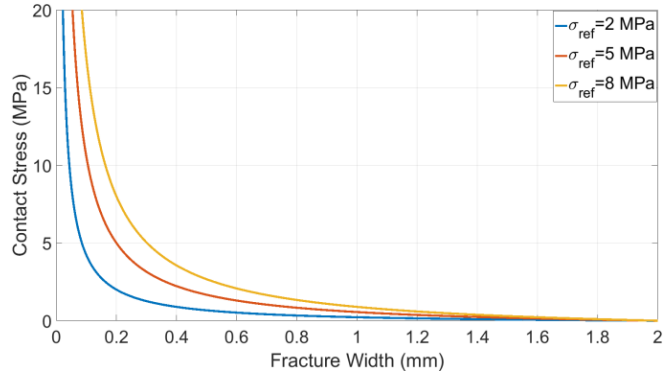

**Fig.8 The relationship between contact stress and fracture width for different $\sigma_{ref}$**

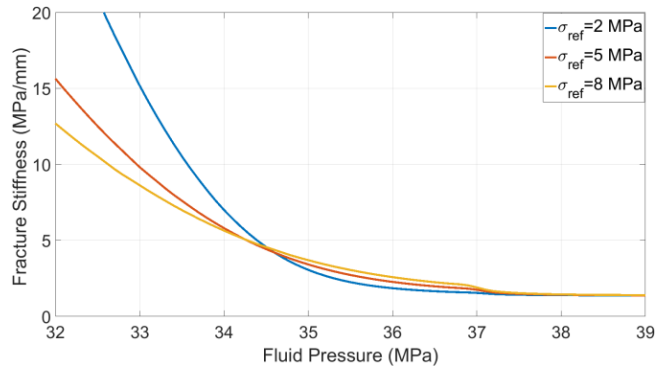

**Fig.9 Fracture stiffness evolution for different $\sigma_{ref}$**

**Fig.10** shows the fracturing pressure and its derivatives for different contact reference stress on G-function and the square root of time plots. Again, we can notice that the contact reference stress has negligible influence on early time pressure decline when fracture stiffness is primarily controlled by initial fracture geometry. However, after this period as the fracture pressure declines further, more and more fracture surfaces come into contact and the contact reference stress begins to affect the pressure decline trend and the peak value of the pressure derivatives.

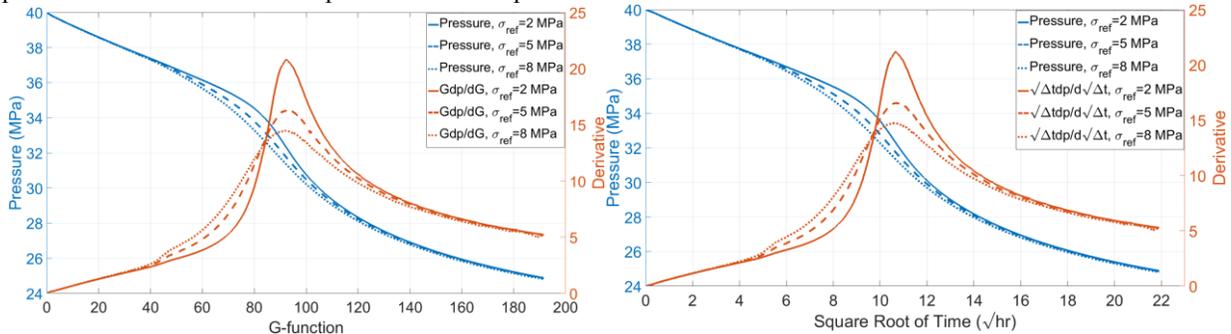

**Fig.10 Pressure decline response for different $\sigma_{ref}$**

**Fig.11** shows the fracture conductivity evolution for different reference contact stress $\sigma_{ref}$. It can be observed that when $\sigma_{ref}$ is small, fracture conductivity is more sensitive to effective stress, this is because $\sigma_{ref}$ determines how supportive the asperities are. When $\sigma_{ref}$ is small, the asperities are more prone to deformation, so the average residual fracture width is smaller at the same level of effective stress.

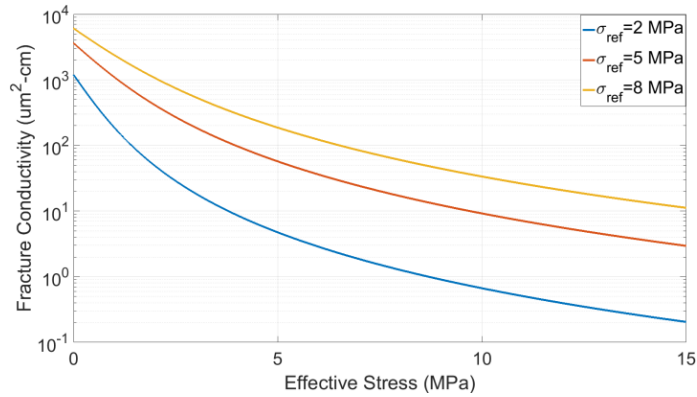

**Fig.11 Fracture conductivity evolution for different $\sigma_{ref}$**

Besides the fracture surface roughness, fracture geometry also affects fracture stiffness and its evolution during closure. **Fig.12** shows fracture stiffness evolution for different fracture height while all the other parameters remain the same as the Base Case. Similar to what Table 1 implies, for a PKN fracture geometry, the smaller the fracture height, the higher the initial fracture stiffness. We can also observe that smaller fracture height leads to noticeable changes in fracture stiffness at higher fracturing pressure. This is because a smaller fracture height results in smaller fracture width at the same net pressure, so more of the fracture surfaces can come into contact at a higher fracture pressure.

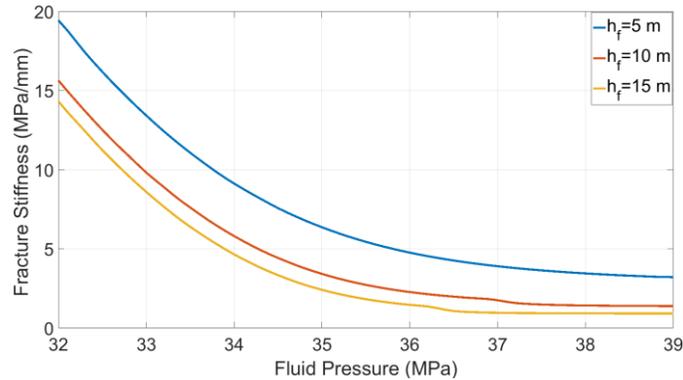

**Fig.12 Fracture stiffness evolution for different fracture height with a PKN geometry**

**Fig.13** shows the fracturing pressure and its derivatives for different fracture height on G-function and square root of time plots. The results indicate that fracture height impacts the pressure decline trend significantly. Larger fracture height leads to later occurrence of the peak of the pressure derivative and this also increases the peak value of the pressure derivatives. The early-time straight lines of pressure derivatives are still controlled by initial fracture geometry.

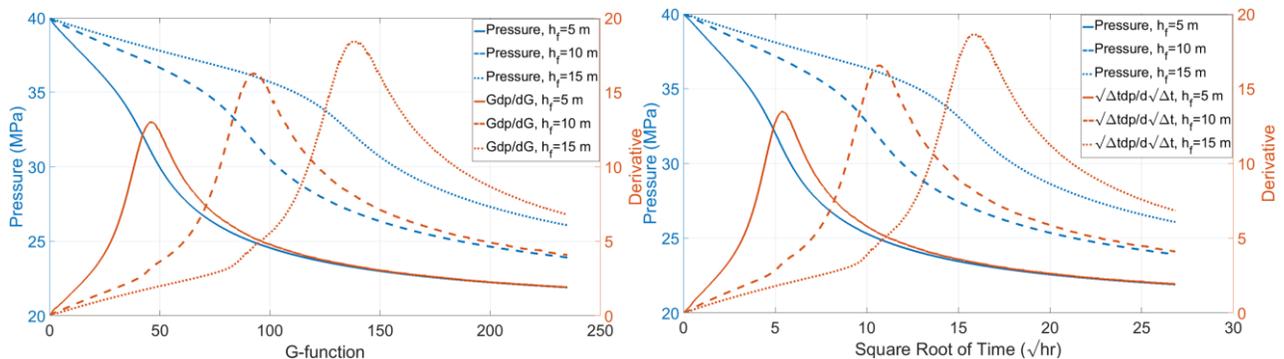

**Fig.13 Pressure decline response for different fracture height with a PKN geometry**

**Fig.14** shows the fracture conductivity evolution for different fracture height with the same fracture roughness (i.e., identical contact parameters). Compared to Fig. 7 and Fig.11, it can be observed that fracture geometry does have an influence on fracture conductivity because it impacts non-local fracture closure process when the effective stress is small (non-uniform contact stress, higher towards the edges), but as the effective stress increases, it is the fracture roughness that dominates the evolution of fracture conductivity as the contact stress becomes more or less non-uniform across fracture surface. So even though fracture geometry may vary, the fracture conductivity is controlled by fracture roughness properties at the end. In other words, if we can obtain representative contact parameters from field data, we can use Eq.(3) to calculate the residual fracture width under a certain stress and correlate the residual fracture width to conductivity via the cubic law.

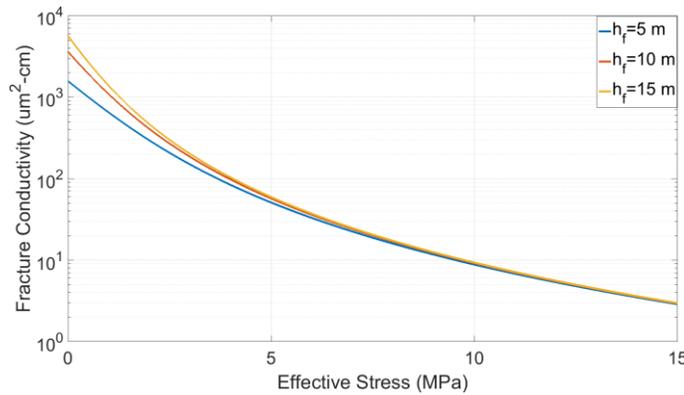

**Fig.14 Fracture conductivity evolution for different fracture height**

**Fig.15** shows the fracturing pressure and its derivatives for different reservoir permeability on G-function and the square root of time plots. As expected, the pressure declines more rapidly when the reservoir permeability is large and the decline rate slows down as the difference between fracturing pressure and initial reservoir pressure becomes smaller.

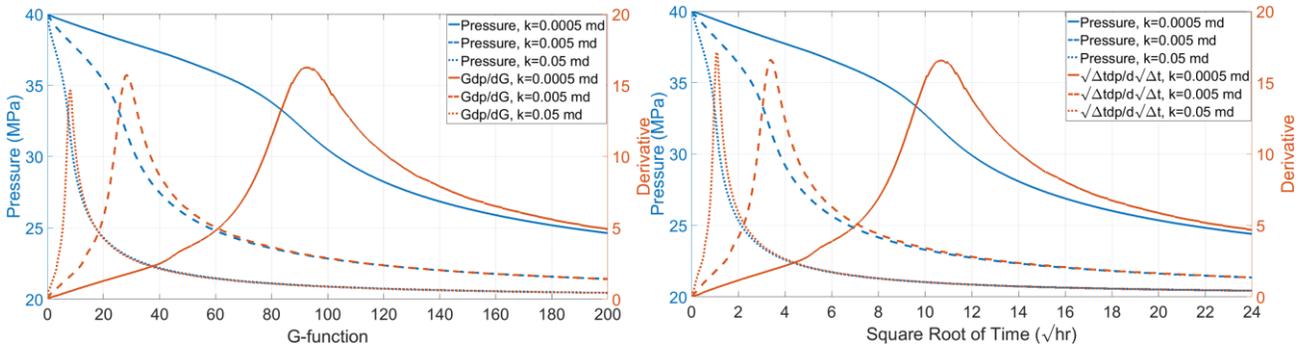

**Fig.15 Pressure decline response for different reservoir permeability with a PKN geometry**

Next, we examine the impact of wellbore storage. The water compressibility is assumed to be 4.35e-4 MPa$^{-1}$. **Fig.16** shows the fracture-wellbore system stiffness evolution for different wellbore volume. As can be seen, when fracturing pressure is high, fracture stiffness dominates the system stiffness. However, as fracturing pressure continues declining, the fracture become less and less compressible and the role of wellbore storage becomes apparent. In general, the larger the wellbore volume, the more gradual and slower the increase in system stiffness will be.

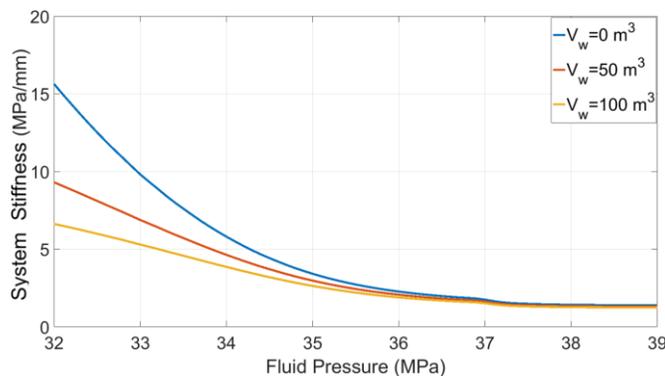

**Fig.16 Fracture-wellbore system stiffness evolution for different wellbore volume with a PKN geometry**

**Fig.17** shows the corresponding fracturing pressure and its derivatives for different wellbore volumes on G-function and square root of time plots. It can be observed that larger wellbore volume leads to more gradual pressure decline trends. A larger wellbore volume also delays the occurrence of fracture closure and lowers the peak of the pressure derivative curve. It can be seen that wellbore storage has a small impact during early time of shut-in when the system stiffness is still dominated by initial fracture stiffness (determined by initial fracture geometry), however, as more and more of the fracture surface comes into contact and the fracture becomes stiffer, wellbore storage effects become apparent, and the after-flow of fluid from wellbore to fracture long after shut-in decelerates the pressure decline rate and extends the tail of the pressure derivative after it reaches the peak.

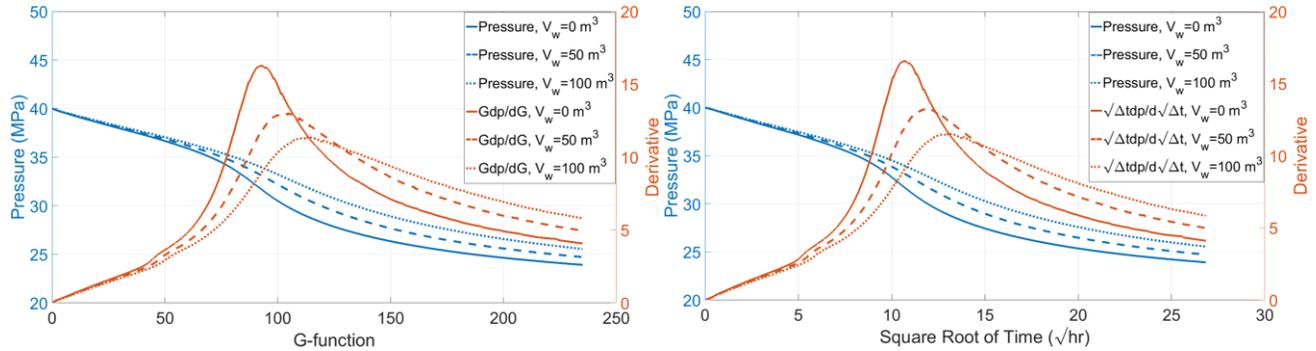

**Fig.17 Pressure decline response for different wellbore volume with a PKN geometry**

**Fig.18** shows the fracturing pressure and its derivatives for different initial reservoir pressure on G-function and square root of time plots. It can be observed when the initial reservoir pressure is low; the pressure declines more rapidly. Because a lower reservoir pressure leads to a higher leak-off rate at the same ISIP. However, when the reservoir is over pressurized with high initial pore pressure, the pressure decline trend resembles a "normal-leak-off behavior". In this case ($P_0$=28MPa), one can still notice that there is a subtle "bump" in the pressure derivatives that indicates an increase in fracture stiffness as reflected in Fig.5. For small values of $\sigma_{ref}$, this increase can be too small and gradual to be noticeable on G-function and square root of time plots. Nevertheless, the early-time pressure decline is still controlled by initial fracture geometry, and the late time pressure derivatives that extend the initial straight line to the peak is coincidence region where the effects of fracture pressure dependent leak-off (leads pressure derivative deviate downward) and the increase of fracture stiffness (leads pressure derivative deviate upward) counterbalance each other.

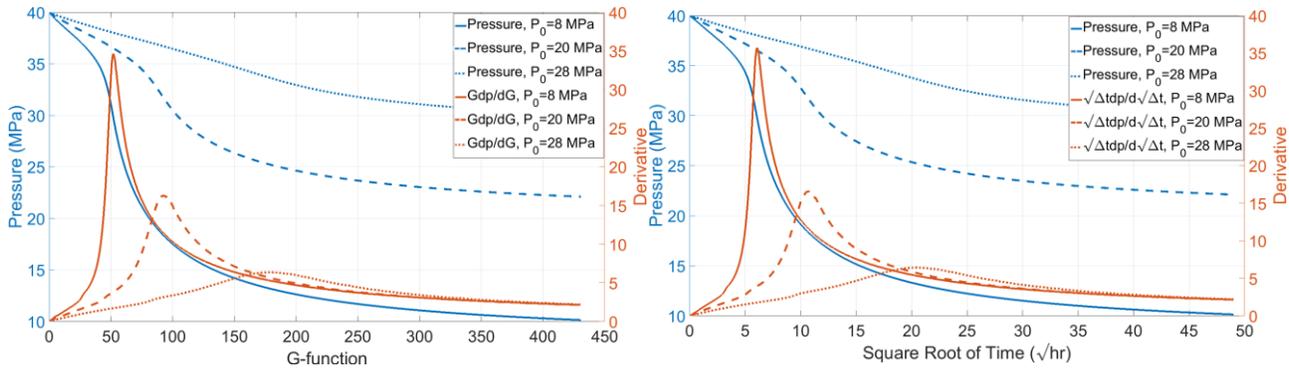

**Fig.18 Pressure decline response for different initial reservoir pressure with a PKN geometry**

From the above synthetic cases and analysis, it is clear that wellbore storage, reservoir properties and fracture stiffness/compliance govern the pressure decline response during DFIT. The fracture stiffness/compliance evolution is determined by rock mechanical properties, fracture geometry and surface roughness (i.e., represented by up-scaled contact parameters). In general, the wellbore storage is known in advance, reservoir properties can be estimated through laboratory experiments, well-logging and after closure analysis. Fracture dimensions with small injection volume can be constrained with proper geological, fracture propagation modeling and early-time stiffness estimation (through Eq.(8) on G-function plot or Eq.(9) on square root of time plot), so it is possible to obtain a good estimate of fracture roughness and infer un-propped fracture conductivity from the analysis of DFIT data. Unlike previous DFIT models, that only focus on one portion of the DFIT data, the DFIT model presented in this study has the capability to model the entire duration (before closure, after closure and the transitional periods), which significantly increases the reliability of data interpretation.

It should be emphasized that the progressive closure for planar fractures will occur primarily in one direction. For example, for a PKN fracture geometry, the fracture closes from top and bottom, because fracture can be considered as "plane strain" in the length direction. For a KGD fracture geometry, the fracture closes from the fracture tip in the length direction, because "plane strain" is assumed in the height direction. And for a radial fracture geometry, the fracture closes radially from the tip of the fracture. For a typical DFIT with a small fluid injection volume, the fracture is mostly likely to be PKN (if fracture height growth is bounded by barrier layers because of a stress contrast) or radial (thick formation layer). The detailed modeling of fracture closure and the difference in fracture height recession, length recession and radius recession are thoroughly discussed in our previous work (Wang and Sharma 2017a; Wang et al. 2018).

# Field Case Studies

## Field Case 1

The first field case analyzed comes from Horizontal Well-A drilled through a shale formation. The measured depth is around 5500 m and a diagnostic fracture injection test is conducted at the toe of the horizontal wellbore, with 2.35 m³ of water injected in 3 minutes, then the well was shut-in for 27 days. **Fig.19** shows the pressure decline response on G-function and the square root of time plots. A closer look at the square root of time plot, shows that the pressure drops significantly during the first one hour of shut-in. By examining the pressure derivative curve, whose extrapolated value is not zero when $\Delta t=0$, we can identify that the excessive pressure drop at the beginning is caused by tip extension and near-wellbore tortuosity. The apparent ISIP is 45.5 MPa while the true ISIP is 34.5 MPa. Remember net pressure is the pressure difference between the fluid pressure inside the fracture and closure pressure. In reality, the pressure measured at the wellhead can be much higher than expected during fracture propagation during the early time of shut-in, because significant pressure drop can happen in the near-wellbore region and along the horizontal wellbore, due to friction and tortuous/complex fracture path that initiated from perforation clusters. So pressure data from this period cannot be used to infer fracturing parameters and reservoir properties.

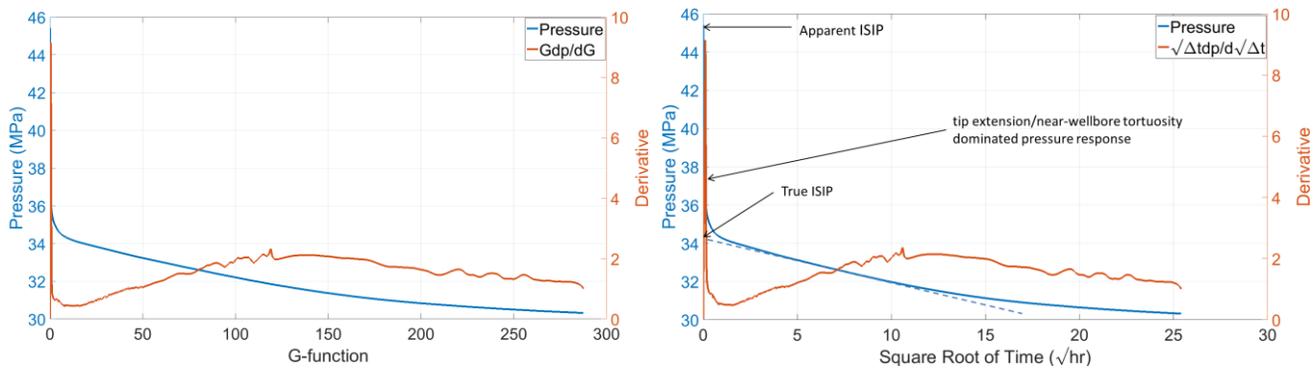

**Fig.19 Pressure decline response from Well-A on G-function and square root of time plots**

From the signature of pressure derivatives, the changes in fracture stiffness/compliance are undetectable, and the minimum in-situ stress is picked at 31.8 MPa from Fig.19. Even though Well-A was shut-in for nearly four weeks, pseudo-radial flow (-1 slope on log-log plot) is still absent, as shown in **Fig.20**. This is a clear evidence of extremely low formation permeability, without the interference of natural fractures.

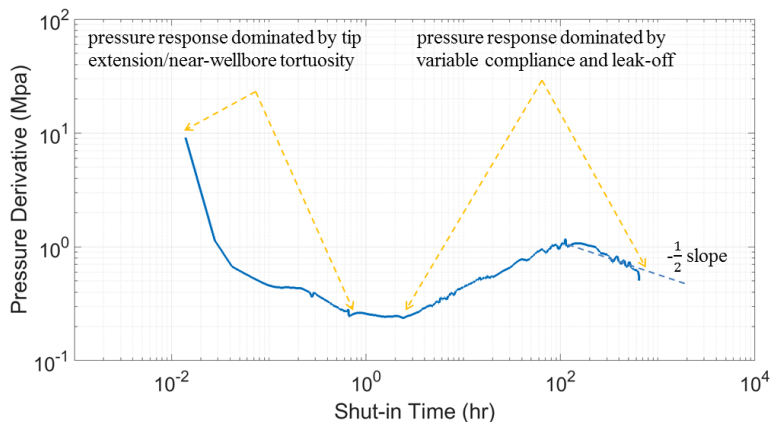

**Fig.20 Log-log pressure derivative plot of Well-A**

Without pseudo-radial flow, formation flow capacity cannot be determined independently with enough confidence using just after closure analysis alone. One has to analyze the whole spectrum of DFIT data to reach a consistent interpretation. Using late time data from the linear flow regime, the estimated initial pore pressure is 29.1 MPa. This indicates it is an over-pressurized reservoir. Before any attempt at detailed modeling and history matching, we first calculate the relative fracture-wellbore stiffness evolution after shut-in using just pressure-time data through Eq.(A12). **Fig.21** shows the normalized fracture-wellbore system stiffness (the ratio $S_s$ at $P_f$ to $S_s$ at ISIP) after shut-in of Well-A, even though the early-time data is impaired by tip extension and friction due to near-wellbore tortuosity, it still provides us qualitative information on the fracture closure process as the system stiffness changes. The results show that after the early-time period (both Fig. 19 and Fig 21, even though generated independently from pressure and time data, show that the early-time pressure abnormality ended around 34 MPa), fracture-wellbore system stiffness gradually increases as fracture pressure declines, until to a point where the system stiffness remains roughly constant. This is the point where the fracture stiffness becomes large enough that its compressibility becomes negligible compared to wellbore storage effect. After the pressure drops below this point, the system storage is essentially equivalent to wellbore storage, this phenomenon is also reflected in Eq.(A10). The most revealing information from Fig.21 is that because fracture closure is a gradual and continuous process from edges to the center, the relative change in stiffness is also a gradual and continuous process. The closure stress lies somewhere between where the system stiffness starts to increase (fracture closure begins at the edges) and where system stiffness reaches a plateau. Even when fracture pressure drops below the closure stress, we still have asperities that support a residual fracture width, which makes a further increase in system stiffness possible (residual fracture width and residual fracture volume can be further reduced) as pressure continues decline (without asperities, the system stiffness reaches a plateau right at the closure stress). It should be emphasized that to produce the normalized system stiffness plot, we do not make any assumptions or need any knowledge of the fracture geometry, fracture surface roughness and reservoir properties, yet the calculated general trend of stiffness evolution agrees very well with the non-local fracture closure modeling and tiltmeter measurement that was shown in Fig.1.

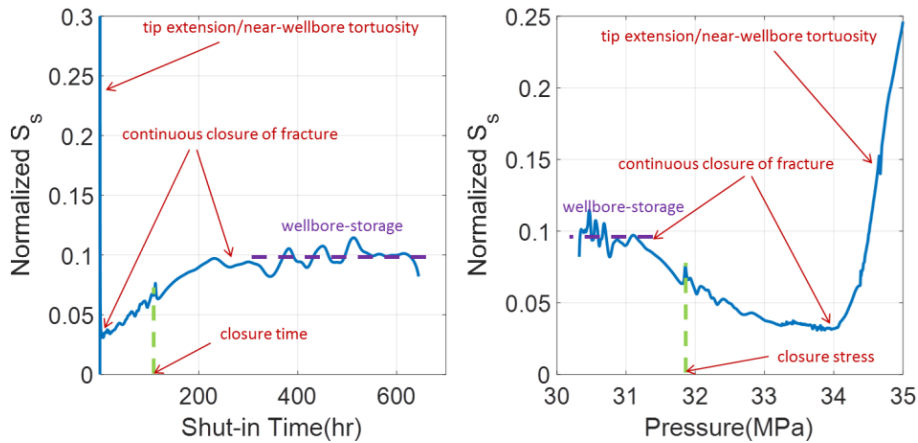

**Fig.21 Normalized fracture-wellbore system stiffness after shut-in of Well-A**

From geological and petrophysical studies, it is known that the thickness of the target formation is 24.4 m with an average porosity of 0.07 and in-situ fluid viscosity of 0.257 cp, The Young's modulus is 38.9 GPa, the Poisson's ratio is 0.2 and formation total compressibility is 3e-3 MPa$^{-1}$. Hydraulic fracture modeling indicates that the fracture is well contained within the target formation with penny-shaped fracture geometry, and the fracture radius is roughly 12 m. Based on this information, the pressure decline response can be matched globally using our DFIT model, from the end of tip extension/near-wellbore tortuosity dominated period to the end of the test.

**Fig.22** shows the pressure decline response predicted by our DFIT model and field data on G-function and the square root of time plots. The results indicate that our simulated pressure matches extremely well with the field data for the entire duration of the test, excluding the first hour of shut-in. Our matched reservoir permeability is 220 nd, which is within the range of independent petrophysical measurements. The matched contact width and contact reference stress are 0.7 mm and 3 MPa, respectively, and **Fig.23** shows the corresponding fracture and fracture-wellbore system stiffness, based on the matched fracture geometry, wellbore volume and contact parameters. Because the DFIT was conducted at the toe of a horizontal well, it is no surprise that the wellbore storage effect is significant, considering such a large contrast between the wellbore and fracture volume. We can also observe that the fracture stiffness has increased 10% when the pressure drops to 33 MPa, over 1 MPa above the true minimum in-situ stress.

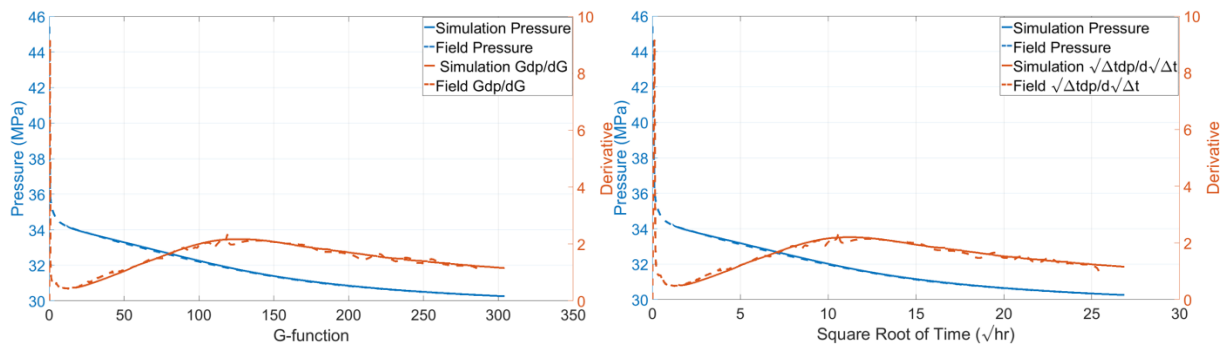

**Fig.22 Matched pressure decline response for Well-A on G-function and square root of time plots**

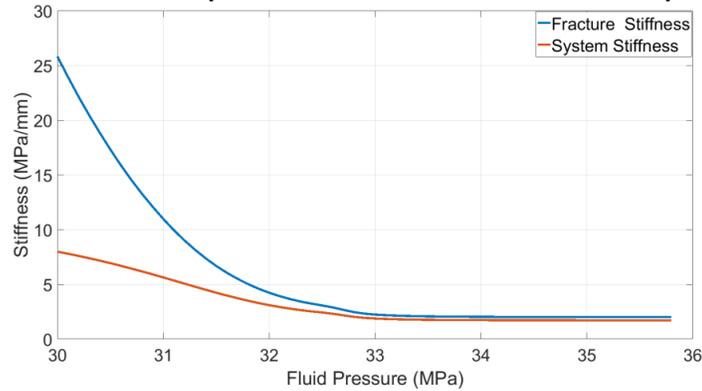

**Fig.23 Matched fracture and fracture-wellbore system stiffness for Well-A**

Similar to traditional pressure transient analysis, a good match of DFIT data does not guarantee that the interpretation is unique. In fact the history matched parameters can be just one set of many combinations that could result in a similar match. So uncertainty analysis can be done to obtain a range of parameters. In the field cases presented here, the fracture dimension has the most uncertainty. When we vary the fracture radius from 10 to 14m, to history match the DFIT data, the matched permeability ranges from 175 nd to 260 nd, the matched contact width ranges from 0.6 mm to 0.8 mm, and the matched contact reference stress ranges from 2.8 MPa to 3.4 MPa. The estimated un-propped fracture conductivity is shown in **Fig.24** for different fracture dimensions. Similar to the previous synthetic cases, fracture conductivity declines more rapidly when the effective stress is small and then follows a semi-log relationship with effective stress as fracturing pressure declines further.

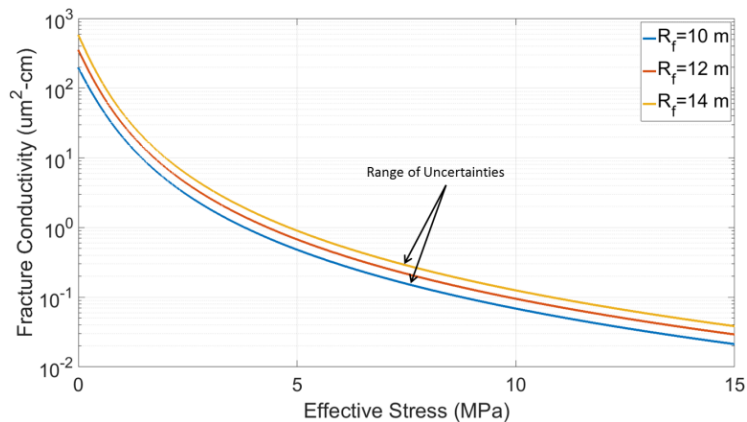

**Fig.24 Estimated range of fracture conductivity and effective stress of an un-propped fracture for Well-A**

## Field Case 2

The second field case to be analyzed comes from a vertical well-B drilled through a shale formation. The total wellbore length is around 2000 m and a diagnostic fracture injection test is conducted with 4.7 m$^3$ of water injection for 6 minutes, then the well was shut-in for 11 days. **Fig.25** shows the pressure decline response on G-function and the square root of time plots.

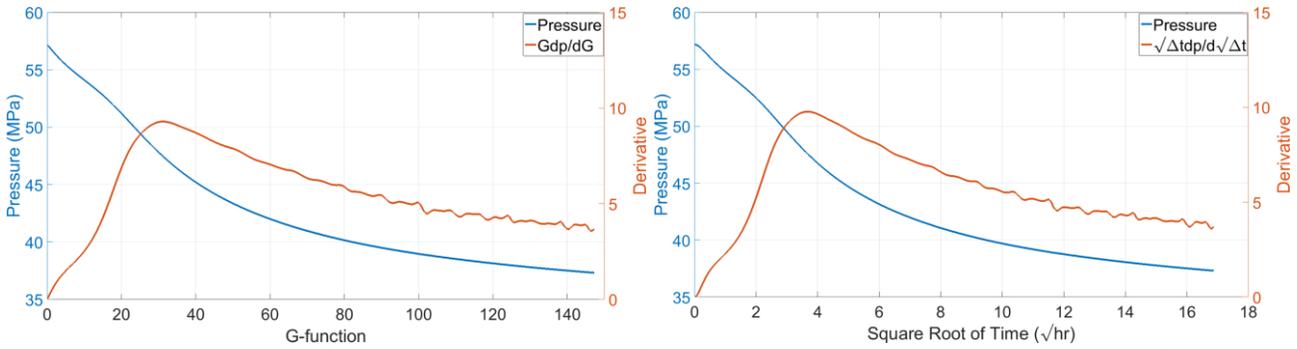

**Fig.25 Pressure decline response from Well-B on G-function and square root of time plots**

Because of the low permeability, no pseudo-radial flow signature can be observed, after-closure linear flow (-1/2 slope on the log-log plot) extends to the end of the test, as shown in **Fig.26**. Using late time data from the linear flow regime, the estimated initial pore pressure is 33.7MPa and the closure stress is picked at 52.4 MPa.

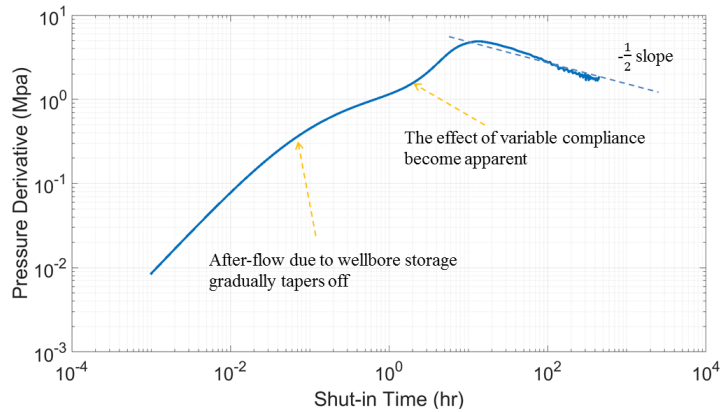

**Fig.26 Log-log pressure derivative plot of Well-B**

Again, before any attempt at detailed modeling and history matching, we first qualitatively examine the fracture-wellbore system stiffness evolution, using just DFIT data and the estimated initial pore pressure. **Fig.27** shows the normalized fracture-wellbore system stiffness after shut-in. Similar to Fig.21, system stiffness gradually increases after early-time abnormality, and eventually reaches a plateau where fracture compressibility becomes negligible. The closure stress lies somewhere in the region where the system stiffness continues to increase smoothly. Together with Fig.21, we can conclude that even though the normalized fracture-wellbore system stiffness plot enables us to extract relative stiffness changes with the least amount of information and gives us the direct evidence of the existence of pressure-dependent residual fracture width, we still need a G-function plot or a square root time plot to estimate closure stress. In other words, because fracture closure is progressive and continuous, closure pressure can't be detected on a normalized stiffness plot or a tiltmeter measurement (which also reflect fracture stiffness evolution), and it has to be picked on the G-function plot or a square root of time plot.

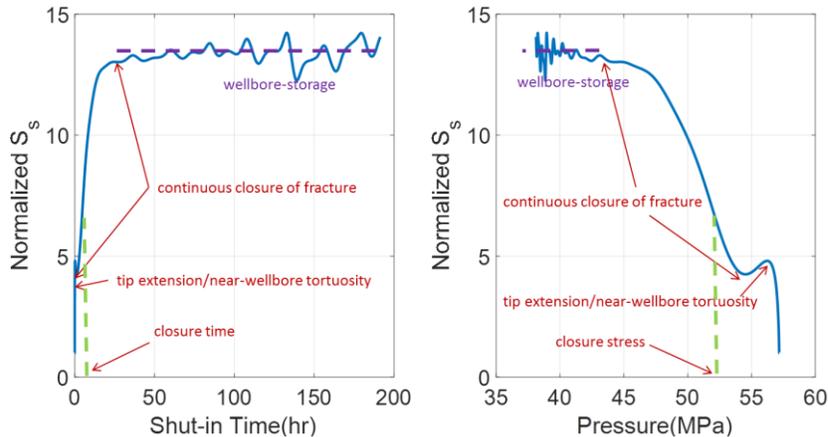

**Fig.27 Normalized fracture-wellbore system stiffness after shut-in of Well-B**

Geological and petrophysical studies indicate that the thickness of the target formation is 5 m with an average effective porosity of 0.03 and an in-situ fluid viscosity of 0.28 cp. The Young's modulus is 39.5 GPa, the Poisson's ratio is 0.25 and formation total compressibility is 1.9e-3 MPa$^{-1}$. Hydraulic fracture modeling shows that the fracture is well contained within the target formation with roughly 200 m fracture half-length. Based on this information and assuming PKN geometry, the pressure decline response can be matched globally using our DFIT model by adjusting reservoir permeability and contact parameters. **Fig.28** shows the predicted pressure decline response and field data on G-function and square root of time plots. Our matched reservoir permeability is 210 nd, and the matched contact width and contact reference stress are 1.2 mm and 1.1 MPa, respectively.

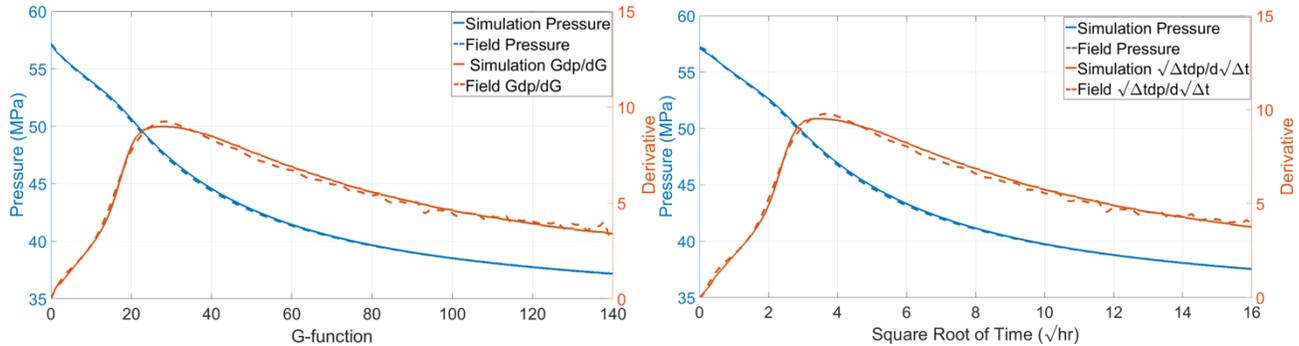

**Fig.28 Matched pressure decline response for Well-B on G-function and square root of time plots**

**Fig.29** shows the corresponding fracture and fracture-wellbore system stiffness as a function of fluid pressure inside the fracture. It can be observed that both fracture and system stiffness increase gradually as pressure declines. Even though this test was conducted in a vertical well with a relatively moderate wellbore volume, the wellbore storage effect on the system stiffness is still significant. This is because the relative influence of wellbore storage depends not only on the ratio of wellbore to fracture volume, but it also on the fracture stiffness itself. For the same wellbore volume, the higher the fracture stiffness, the less compliant the fracture is and the compressed fluid in the wellbore plays a bigger role. In this field case, with fracture height is contained in 5 m, the initial fracture stiffness should be around 5.36 MPa/mm (calculated from Table 1).

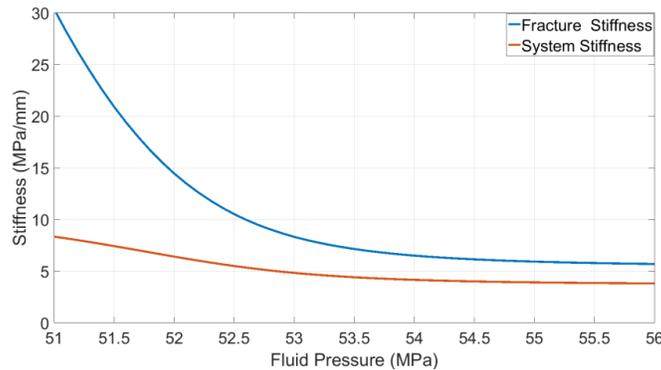

**Fig.29 Matched fracture and fracture-wellbore system stiffness for Well-B**

Even though we know that the fracture is bounded with a PKN type fracture geometry, there are still some uncertainties in fracture height. We vary the fracture height from 3 m to 7 m, and history match DFIT data, the matched permeability ranges from 124 nd to 300 nd, the matched contact width ranges from 0.8 mm to 1.5 mm, and the matched contact strength ranges from 0.9 MPa to 1.3 MPa. The ranges of estimated un-propped fracture conductivity are shown in **Fig. 30**. Compared to Fig.24 (Well-A), we notice that the un-propped fracture conductivity of Well-B is more sensitive to effective stress, this is because the reference contact stress $\sigma_{ref}$ of Well-B is smaller than that of Well-A, so the asperities are more prone to deformation at the same effective stress, resulting in a smaller residual fracture width.

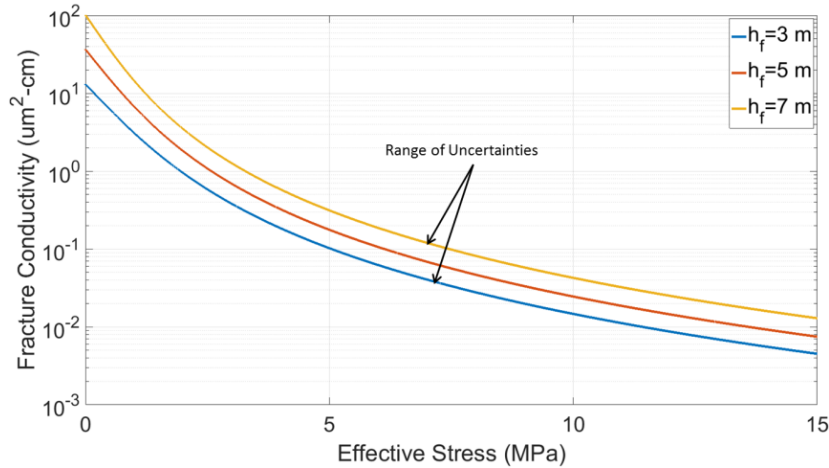

**Fig.30 Estimated range of fracture conductivity and effective stress of un-propped fracture for Well-B.**

In fact, if we compare Fig. 21 and Fig.27. the difference between the closure stress and the start of the stiffness plateau is 0.8 MPa for Well A and 8 MPa for Well B, this means that the residual fracture width is more compliant and more prone to deformation in Well B than in Well A. Consequently, the un-propped fracture conductivity should be less sensitive to an increase in the effective stress in Well A. This agrees well with our modeling results ($\sigma_{ref}$ of Well-A > $\sigma_{ref}$ of Well-B). If we examine the difference between closure stress and where system stiffness starts to increase noticeably, it is around 1.2 MPa for Well-A and 1.6 MPa for Well-B. This difference is largely controlled by the height of fracture surface asperities. Higher surface asperities lead to a larger difference because the fracture can close on asperities at a higher pressure at the fracture edges. So the fracture in Well-B should have taller representative surface asperities. This also agrees well with our modeling results ($w_0$ of Well-B > $w_0$ of Well-A).

It should be emphasized that the forward modeling and history matching of the DFIT data is completely independent of the analysis of the normalized system stiffness plot, yet the two analyses are incredibly consistent. In other words, by just comparing the pressure difference between the start of the plateau of system stiffness and the closure stress on a normalized system stiffness plot, we can infer qualitatively how deformable the asperities are and how resilient the un-propped fracture conductivity is to stress. Also, by comparing the pressure difference between closure stress and the point where system stiffness increases noticeably, we can infer the relative height of the fracture surface asperities. Ideally, to maintain high un-propped fracture conductivity, we would like to have a rough fracture surface that has tall surface asperities and the residual fracture should be as stiff as possible. These characteristics can be inferred from the normalized system stiffness plot as: 1) a large difference between the closure stress and the beginning of the increase in system stiffness, 2) a small difference between closure stress and the start of the system stiffness plateau. So even without doing detailed forward modeling and history matching of DFIT data, the normalized system stiffness plot using just the pressure and time data can give us valuable information on the evolution of fracture conductivity and stiffness as the effective stress increases.

## Conclusions

A new method is presented that allows us to model pressure transient behavior of a closing fracture and estimate fracture compliance (or stiffness) and unpropped fracture conductivity from DFIT data. Fracture compliance and conductivity can be measured in a laboratory using core samples, however, in many cases, preserved core samples are not available, and most importantly, these measurement results are highly dependent on how the fracture is created under laboratory conditions on specific samples and may not be presentative of a field scale fracture because of reservoir heterogeneity and thin laminated rock layers.

Since the introduction of Nolte's pioneering work using the G-function approach, diagnostic fracture injection tests (DFIT) have been generally accepted as a reliable way to obtain fracturing parameters such as minimum in-situ stress, leak-off behavior and reservoir properties (e.g., initial pore pressure and representative formation permeability). These key parameters are needed to run fracture models, post-fracture production prediction and economic evaluations. However, G-function models and subsequent related studies assume that the fracture compliance remains constant and Carter's leak-off (leak-off rate solution derived from a constant pressure boundary condition) during fracture closure. This prevents these models from predicting the pressure decline response correctly.

In this article, we have proposed a new semi-analytical DFIT model which has the capability to predict the pressure decline response across the entire duration of a DFIT test based on time-convolution. With forward modeling, we show how each factor impacts the pressure decline response. Results indicate that it is possible to estimate surface roughness and infer un-propped fracture conductivity from history matching of DFIT data. Most importantly, the new DFIT model enables us to estimate these quantities using just pressure and time DFIT data. The normalized system stiffness plot (proposed in this paper) allows us to accomplish this without making any assumptions or any knowledge of fracture geometry, fracture surface roughness and reservoir properties. This specialized plot extracts and highlights information on the progressive fracture closure behavior and the properties of fracture surface roughness, which can be used to infer un-propped fracture conductivity qualitatively. The fracture stiffness evolution during closure revealed by the normalized system stiffness plot agrees well with non-local fracture closure modeling and tiltmeter measurements. Field cases are presented that also indicate that the analysis from the normalized system stiffness plot is consistent with the quantitative analysis of our forward modeling and history match.

Because both the fracture and wellbore storage impact the system stiffness in the field, it is preferable to design a diagnostic fracture injection test where the wellbore storage effect can be minimized (i.e., using downhole shut-in), so the influence of fracture properties are more apparent on diagnostic plots. There are some limitations on when our model can be applied: 1) it only applies to planar fractures and cannot be used for complex fractures where multiple peaks in the pressure derivative plot are observed. 2) The reservoir permeability is assumed to not be strongly pressure-dependent within the fall-off period, otherwise it can partly mask the influence of fracture closure. In other words, if only one peak is observed in the pressure derivative on G-function or square root of time plot without a "hump (pressure dependent permeability signature)" before closure, together with a long period of after-closure linear flow (further confirmation of planar fracture geometry), it is a good indication that forward modeling and history matching can be confidently applied to estimate the properties of the un-propped fracture.

However, there are fewer restrictions on the use of the normalized system stiffness plot, as long as leak-off is linear and the reservoir permeability is not strongly pressure dependent within the fall-off pressure range, the normalized system stiffness plot provides new information about the evolution of the fracture stiffness with effective stress, regardless of whether the fracture geometry is planar or complex. This, in turn, provides us a unique insight into the dynamic closure behavior of unpropped fractures. We recommend that the proposed normalized system stiffness plot be standard practice to complement the G-function or square root of time plot, because it provides valuable information on field-scale fracture surface roughness, fracture compliance evolution and stress-dependent conductivity that cannot be obtained by any other means. This adds tremendous value to DFIT data.

## Nomenclature

| | |
|---|---|
| $A_f$ | = Fracture surface area (one face of one wing), $L^2, m^2$ |
| $c_t$ | = Formation total compressibility, $Lt^2/m$, 1/Pa |
| $c_w$ | = Water compressibility, $Lt^2/m$, 1/Pa |
| $C_L$ | = Carter's leak-off coefficient, $L/\sqrt{t}, m/\sqrt{s}$ |
| $C_s$ | = Fracture-Wellbore storage coefficient, $L^4t^2/m$, m$^3$/Pa |
| $E$ | = Young's modulus, $m/Lt^2$, $Pa$ |
| $E'$ | = Plane strain Young's modulus, $m/Lt^2$, $Pa$ |
| $G(\Delta t_D)$ | = Dimensionless G-function of time |
| $h_f$ | = Fracture height, $L, m$ |
| $ISIP$ | = Instant shut-in pressure, $m/Lt^2$, $Pa$ |
| $k$ | = Formation permeability, $L^2, m^2$ |
| $P$ | = Pressure, $m/Lt^2$, $Pa$ |
| $P_f$ | = Fracturing pressure, $m/Lt^2$, $Pa$ |
| $P_{net}$ | = Fracturing net pressure, $m/Lt^2$, $Pa$ |
| $P_0$ | = Initial reservoir pressure, $m/Lt^2$, $Pa$ |
| $q_f$ | = Leak-off rate (one wing), $L^3/t, m^3/s$ |
| $r_p$ | = Productive surface ratio, which is the ratio of fracture surface area that is subject to leak-off to the total fracture surface area. |
| $R_f$ | = Fracture radius, $L, m$ |
| $S_f$ | = Fracture stiffness, which is the reciprocal of fracture compliance, $m/L^2t^2, Pa/m$ |
| $S_s$ | = Fracture-wellbore system stiffness, $m/L^2t^2, Pa/m$ |
| $t$ | = Generic time, $t, s$ |
| $\Delta t$ | = Total shut-in time, $t, s$ |
| $x_f$ | = Fracture half-length, $L, m$ |
| $U_f$ | = Leak-off velocity, $L/t, m/s$ |
| $V_f$ | = Fracture volume (one wing), $L^3, m^3$ |
| $V_w$ | = Half of wellbore volume, $L^3, m^3$ |
| $w_0$ | =Contact width, $L, m$ |
| $w_f$ | =Local fracture width, $L, m$ |

$\bar{w}_f$ = Average fracture width, L, $m$
$\mu_f$ = Fluid viscosity, m/Lt, Pa·s
$\nu$ = Poisson's ratio
$\sigma_c$ = Contact stress, m/L$t^2$, $Pa$
$\sigma_{hmin}$ = Minimum in-situ stress, m/L$t^2$, $Pa$
$\sigma_{ref}$ = Contact reference stress, m/L$t^2$, $Pa$
$\phi$ = Formation porosity

## Appendix: A Semi-analytical DFIT Model with Variable Fracture Compliance and Fracture Pressure Dependent Leak-off

The pressure transient response during fracture closure is derived using the following assumptions:

1. Reservoir is isotropic and homogeneous and contains a single slightly compressible fluid, and the injected fluid has the same properties as the reservoir fluid. This assumption is appropriate as long as the PVT properties used in the DFIT model represent in-situ fluid (Gu et al. 1993; Soliman et al. 2005).
2. The fluid viscosity, formation porosity, total compressibility, and rock matrix permeability are independent of pressure.
3. Reservoir permeability is low so that poroelastic effects caused by fluid leak-off are negligible
4. Gravity effects are negligible.
5. Leak-off surface area is constant. This means that mechanically closed fracture still retains hydraulic conductivity because of its residual fracture width that supported by asperities that caused by erosion or distortion of fracture walls.
6. The pressure is uniformly distributed inside the fracture. This is the typical case in unconventional reservoirs. The pressure distribution inside fracture can be considered to be uniform after shut-in and during closure, as discussed by Koning et al. (1985). The existence of after-closure linear flow also demonstrates that fracture conductivity can be regarded as infinite when the pressure drop along the fracture is negligible.
7. The pore pressure disturbance caused by fracture propagation is negligible. This assumption is reasonable because fluid leak-off during pumping is small and the duration of injection is short (typically 3-10 minutes) while the total shut-in time can be hours, days or even weeks.
8. Leak-off is linearly perpendicular to fracture surface and late time radial flow has not been developed yet.

In order to correctly capture fracturing pressure response during a DFIT, the pressure-dependent leak-off at the fracture surface and the dynamic changes of fracture compliance during closure have to be accounted for. **Fig.A1** illustrates one-dimensional leak-off into a semi-infinite formation.

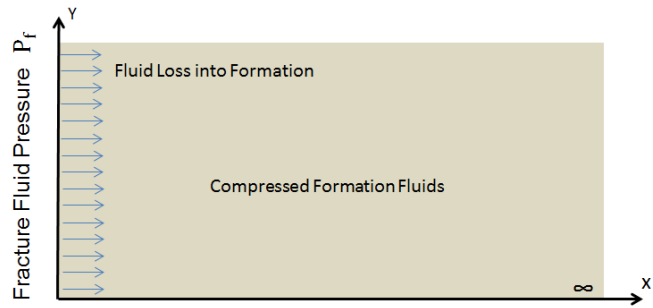

**Fig. A1 Illustration of one-dimensional leak-off**

Assuming linear Darcy flow and a slightly compressible, single phase fluid in the reservoir, the differential form of the mass balance can be written as:

$$\frac{\mu_f \phi c_t}{k} \frac{\partial P}{\partial t} = \frac{\partial^2 P}{\partial x^2} \tag{A1}$$

where $P$ is the pressure, k is formation permeability, $\mu_f$ is fluid viscosity, $\phi$ is formation porosity and $c_t$ is total formation compressibility. If we assume that a constant fracture pressure $P_f$ is applied at the fracture surface, the leak-off velocity $U_f$ across fracture surface can be found as (Economides and Nolte 2000):

$$U_f = (P_f - P_0)\sqrt{\frac{k\phi c_t}{\pi \mu_f \Delta t}} \tag{A2}$$

where $P_0$ is the initial reservoir pressure. The total leak-off rate $q_f$ from one wing of the fracture starting from shut-in is

$$q_f = 2A_f(P_f - P_0)\sqrt{\frac{k\phi c_t}{\pi \mu_f \Delta t}} \tag{A3}$$

where $A_f$ is the fracture surface area of one face of one wing of the fracture. If $P_f$ is constant, then the leak-off rate $q_f$ is proportional to $\frac{1}{\sqrt{\Delta t}}$, which is the assumption of Carter's leak-off model. However, $P_f$ continues decline during fracture closure, the leak-off rate will deviate from Carter's leak-off model and turn the pressure derivative downward on G-function and square root of time plots (Wang and Sharma 2017b). To account for the fracture pressure dependent leak-off, pressure superposition is needed. Divide the shut-in time $\Delta t$ into n time steps, and the leak-off rate at the n[th] time step can be determined based on superposition:

$$q_{f,n} = 2A_f\sqrt{\frac{k\phi c_t}{\pi \mu_f}} \sum_{j=1}^{n} \frac{P_{f,j} - P_{f,j-1}}{\sqrt{\Delta t_n - \Delta t_{j-1}}} \tag{A4}$$

The pressure difference $P_{f,j} - P_{f,j-1}$ is the pressure in the fracture at time step $j$ minus the pressure in the formation at the fracture-formation interface, which equals to the pressure in the fracture at the previous time step. From a material balance perspective (fluid compressibility is negligible compared to that of the fracture), the rate of fluid leak-off into the formation, $q_f$ (one wing of the fracture), after shut-in equals the rate of shrinkage of fracture volume, $V_f$ (one wing of the fracture), as pressure declines:

$$q_f = -\frac{dV_f}{dt} \tag{A5}$$

The fracture stiffness can be obtained from fracture closure modeling (Wang and Sharma 2017a; Wang et al. 2018) using:

$$S_f = \frac{A_f dP_f}{dV_f} \tag{A6}$$

For open fractures at relative high pressure, $S_f$ can be determined analytically using Table-1. For more general circumstances, $S_f$ can be calculated numerically as the fracture closes progressively (Wang and Sharma 2017a; Wang et al. 2018), and the numerical solution gives the same results as using the formulae in Table-1 when the fracture is still open at high pressure.

With the above definitions, Eq.(A5) can be re-written as

$$q_f = -\frac{A_f}{S_f}\frac{dP_f}{dt} \tag{A7}$$

If we discretize Eq.(A7) into small time intervals where the pressure drop is insignificant within each interval, then the term $\frac{A_f}{S_f}$ can be treated as a constant in each time interval, though the value of $\frac{A_f}{S_f}$ is pressure-dependent for different time intervals. Discretize Eq.(7) into $q_{f,n} = -\frac{A_f}{S_{f,n}}\frac{dP_{f,n}}{dt_n}$ and equate it to Eq.(A4), the $A_f$ is then canceled out in each time interval:

$$\frac{1}{S_{f,n}}\frac{dP_{f,n}}{d\Delta t_n} = -2\sqrt{\frac{k\phi c_t}{\pi \mu_f}} \sum_{j=1}^{n} \frac{P_{f,j} - P_{f,j-1}}{\sqrt{\Delta t_n - \Delta t_{j-1}}} \tag{A8}$$

The R.H.S. of Eq.(A8) comes from the solution of pressure-dependent leak-off rate in Eq.(A4). The L.H.S. of Eq.(A8) comes from the discretization of material balance in Eq.(A7), so superposition is only used in Eq.(A4). Eq.(A8) is implicit, and to make superposition explicit and obtain time-convolution solution, we need to make additional assumptions for discretization (similar to the derivation of time-convolution solution in heat transfer problem by Zhang et al. 2011 ): the time interval is small, the fracture stiffness is constant during current time step and equals the value at the beginning of current time step. Adjust the index so that $P_{f,0} = P_0$, $P_{f,1} = ISIP$, $\Delta t_1 = 0$, and then for n $\geq$ 2:

$$\frac{dP_{f,n}}{d\Delta t_n} = -2S_{f,n-1}\sqrt{\frac{k\phi c_t}{\pi \mu_f}} \sum_{j=1}^{n-1} \frac{P_{f,j} - P_{f,j-1}}{\sqrt{\Delta t_n - \Delta t_{j-1}}} \tag{A9}$$

Integrate Eq.(A9) across the discretized data points with respect to shut-in time, we can obtain fracturing pressure with changing fracture stiffness and fracture pressure dependent leak-off based on a time-convolution solution :

$$P_{f,n} = ISIP - 4\sqrt{\frac{k\phi c_t}{\pi \mu_f}} \sum_{i=1}^{n-1} S_{f,i} \sum_{j=1}^{i} (P_{f,j} - P_{f,j-1})\left(\sqrt{\Delta t_i - \Delta t_{j-1}} - \sqrt{\Delta t_{i-1} - \Delta t_{j-1}}\right) \tag{A10}$$

From Eq.(A10), we realize that for a given initial condition, reservoir properties and pressure dependent fracture stiffness, the pressure decline response is uniquely determined. In the above derivation, it is assumed that the whole fracture surface area is subject to leak-off, which is the norm in unconventional reservoirs. If only a portion of fracture surface is considered permeable, then one needs to multiply the right side of Eq.(A4) by the productive fracture ratio $r_p$. To account for wellbore storage effects, the fracture stiffness $S_f$ in Eq.(A10) needs to be replaced by the fracture-wellbore system stiffness, which is defined as:

$$S_s = \frac{A_f}{\frac{A_f}{S_f} + V_w \, c_w} = \frac{S_f A_f}{A_f + S_f V_w \, c_w} \tag{A11}$$

Where $V_w$ is half the wellbore volume (only one wing of the fracture needs to be modeled) and $c_w$ is the compressibility of water. In essence, the fracture-wellbore system stiffness reflects the fracture surface area normalized system compressibility. From Eq.(A11), we can infer that when fracture stiffness is small (fracture compliance is large), the system stiffness is dominated by fracture stiffness, because the fracture has a large compressibility compared to the wellbore. However, as fracture stiffness continues to increase during closure, the wellbore storage will play a more and more important role, and at the end, when fracture stiffness becomes large enough, the system stiffness is controlled entirely by wellbore storage, and becomes independent of pressure. In other literature, the system compressibility sometimes can be referred to as system storage coefficient $C_s$:

$$C_s = \frac{A_f}{S_s} = \frac{A_f}{S_f} + V_w \, c_w \tag{A12}$$

**Fig.A2** compares the results from the time-convolution solution (with a discretized time interval of 5 s) with the numerical model presented by (Wang and Sharma 2017b) for the Base Case scenario, it demonstrates that the time-convolution gives exactly the same solution as the numerical simulation.

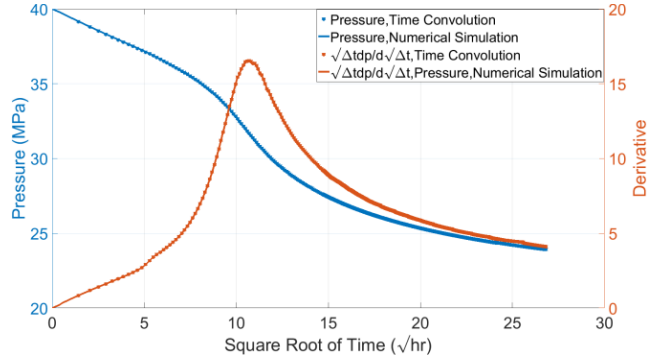

**Fig.A2 Comparison of the time convolution solution against a numerical simulation**

Compared with the numerical model (Wang and Sharma 2017b) for solving the coupled PDE-ODE system, the time convolution solution is not only more computationally efficient, but can also give us some qualitative information on the evolution of fracture or system stiffness using just pressure and time data alone. Let us assume that $4\sqrt{\frac{k\phi c_t}{\pi \mu_f}}$ in Eq(A10) can be represented by a constant number $\Psi$, then Eq.(A10) can be rearranged to solve for fracture-wellbore system stiffness as:

$$S_{s,n-1} = \frac{\frac{ISIP - P_{f,n}}{\Psi} - \sum_{i=1}^{n-2} S_{s,i} \sum_{j=1}^{i} (P_{f,j} - P_{f,j-1})(\sqrt{\Delta t_i - \Delta t_{j-1}} - \sqrt{\Delta t_{i-1} - \Delta t_{j-1}})}{\sum_{j=1}^{n-1} (P_{f,j} - P_{f,j-1})(\sqrt{\Delta t_i - \Delta t_{j-1}} - \sqrt{\Delta t_{i-1} - \Delta t_{j-1}})} \tag{A13}$$

Using this formulation, we can now get the evolution of normalized stiffness $S_{s,n-1}/S_{s,1}$ just using just the DFIT pressure and time data (with initial pore pressure estimated through after-closure analysis), and the results are independent of the value of $\Psi$. For example, **Fig.A3** shows early-time DFIT data on G-function and square root of time plots that resembles a perfect "normal leak-off" behavior, as indicated by straight lines of pressure derivatives. Assume from after-closure analysis we get the initial pore pressure $P_{f,0}$, then based on Eq.(A13) we can calculate the normalized fracture-wellbore system stiffness evolution using only shut-in time $\Delta t_i$ and corresponding pressure $P_{f,i}$ by assigning a random number to $\Psi$. The value of $\Psi$ will impact the calculated value of $S_{s,i}$, but it will not change the value of $S_{s,i}/S_{s,1}$. The calculated system stiffness using the simulated shut-in time and pressure is shown in **Fig.A4**. The results clearly reveal that the fracture stiffness gradually increases during closure. In addition, we can notice that the initial pore pressure impact the absolute value of the normalized system stiffness calculation, but does not alter the general trend of normalized system stiffness evolution. This means that in rare cases, even if we cannot obtain the initial pore pressure from after-closure analysis, by assigning the initial pore pressure a reasonable value, we can still produce the normalized system stiffness plot and estimate fracture surface roughness properties. This is possible because the analysis relies on the deflection points of normalized system stiffness and the corresponding pressure, not its absolute value. Ideally, the normalized system stiffness should start from one and increase gradually as the fracture closes progressively, but in reality, early-time data is often impaired by tip extension and friction due to near-wellbore tortuosity. Under such a scenario, $S_{s,1}$ does not reflect the true system stiffness at ISIP and the scale of $S_{s,i}/S_{s,1}$ will be distorted. Nevertheless, the general trend of normalized system stiffness after the early-time abnormality still provides us qualitative information on the progressive fracture closure behavior.

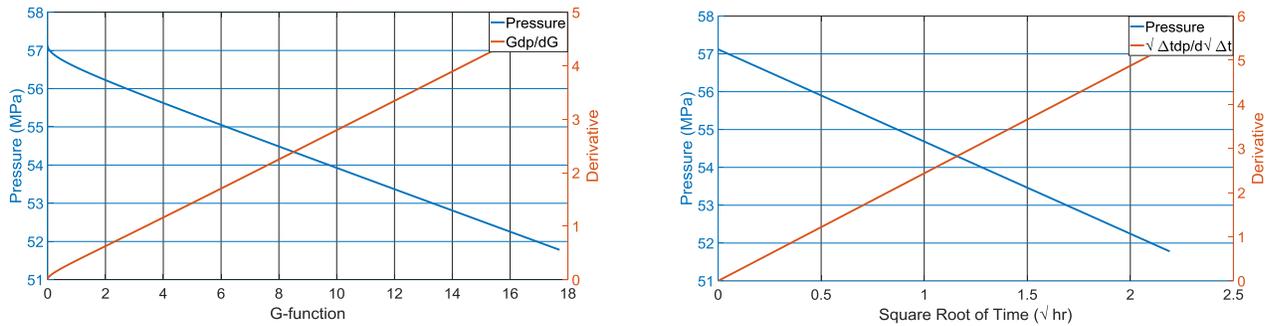

**Fig.A3 Example of "normal leak-off" behavior on G-function and square root of time plots**

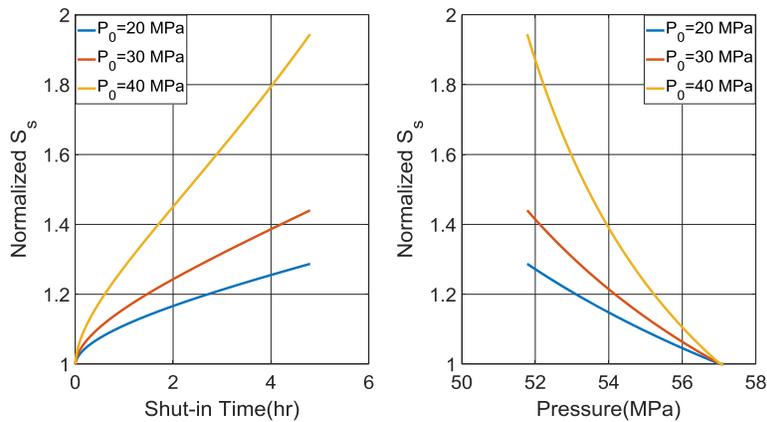

**Fig.A4 The evolution of normalized fracture-wellbore system stiffness after shut-in for different initial pore pressure with "normal leak-off" behavior**

To better familiarize the readers with our proposed method, we provide a step by step guideline to perform our new analysis of DFIT data:

1) Construct G-function or square root of time plot and pick the true ISIP and closure stress. If a pressure-dependent permeability signature (a "bump" occurs on pressure derivative curve) dominates the entire before-closure data, it is an indication that the formation is heavily naturally fractured within the distance of investigation and our approach is not applicable in such a case.

2) Plot $\Delta t \frac{d\Delta P_f}{d\Delta t}$ vs $\Delta t$ on a log-log scale to identify flow regimes.

3) Estimate initial pore pressure from after-closure analysis.

4) Create a normalized system stiffness plot using estimated initial pore pressure, and DFIT pressure-time.

5) Mark closure stress and closure time on a normalized system stiffness plot and evaluate the properties of fracture surface roughness qualitatively.

6) Estimate and constrain fracture geometry and dimensions using other independent information, such as injection volume, formation thickness, stress contrast, etc.

7) Obtain in-situ fluid PTV properties and rock mechanical properties.

8) Generate a fracture stiffness vs. pressure curve through numerical modeling (Wang and Sharma 2017; Wang et al. 2018), and calculate the fracture-wellbore system stiffness using Eq.(A11)

9) History match DFIT data using Eq.(A10) to the end of after-closure linear flow by adjusting two contact parameters (i.e., $w_0$ and $\sigma_{ref}$ in step 8) and formation permeability

10) Generate un-propped fracture conductivity vs. effective stress using matched contact parameters through cubic law

for different residual fracture width.

11) Repeat above steps from 8) to 10) over a range of fracture geometries (dimensions that are constrained within possible range). Tabulate fracture geometry (length) vs. matched formation permeability and un-propped fracture conductivity to account for uncertainties. The uncertainty analysis on other parameters can also be performed if necessary.

12) If after-closure radial flow occurs, after-closure radial flow analysis can be performed independently to further refine fracture geometry and corresponding formation permeability and un-propped fracture conductivity.